# Quantifying the charge carrier interaction in metallic twisted graphene superlattices

Evgueni F. Talantsev[1,2]

[1]M.N. Mikheev Institute of Metal Physics, Ural Branch, Russian Academy of Sciences, 18, S. Kovalevskoy St., Ekaterinburg, 620108, Russia

[2]NANOTECH Centre, Ural Federal University, 19 Mira St., Ekaterinburg, 620002, Russia

## Abstract

The mechanism of charge carrier interaction in twisted bilayer graphene (TBG) remains an unresolved problem, where some researchers proposed the dominance of the electron-phonon interaction, while the others showed evidence for electron-electron or electron-magnon interactions. Here we propose to resolve this problem by generalizing the Bloch-Grüneisen equation and using it for the analysis of the temperature dependent resistivity in TBG. It is a well-established theoretical result that the Bloch-Grüneisen equation power-law exponent, $n$, exhibits exact integer values for certain mechanisms. For instance, $n$=5 implies the electron-phonon interaction, $n$=3 is associated with electron-magnon interaction and $n$=2 applies to the electron-electron interaction. Here we interpret the linear temperature-dependent resistance, widely observed in TBG, as $n \approx 1$, which implies the quasielastic charge interaction with acoustic phonons. Thus, we fitted TBG resistance curves to the Bloch-Grüneisen equation, where we propose that $n$ is a free-fitting parameter. We found that TBGs have a smoothly varied $n$-value (ranging from 1.4 to 4.4) depending on the Moiré superlattice constant, $\lambda$, or the charge carrier concentration. This implies that different mechanisms smoothly transition from one to another. The proposed generalized Bloch-Grüneisen equation is applicable to a wide range of problems, including the Earth geology.

**Quantifying the charge carrier interaction in metallic twisted graphene superlattices**

Bilayer graphene with twisted atomic sheets represents a versatile two-dimensional material where depends on the rotation angle and change carrier concentration a wide number of physical effects can be emerged [1-11]. One of the most interesting subclass of twisted bilayer graphene structures is the Moiré superlatices formed at so-called magic angles, θ, at which two layers become more strongly coupled and the Dirac velocity crosses zero [12-14]. TBG hexagonal two-dimensional superstructures are characterised by the superlattice constant, λ [13]:

$$\lambda = \frac{a}{2 \cdot sin(\theta)} \qquad (1)$$

where $a$ = 0.246 nm is the lattice constant of the single layer graphene. It should be noted that recently Park *et al* [15] reported that magic-angle trilayer graphene superlattices have correlated electronic states similar to ones observed in its bilayer counterpart.

One of most interesting problem in understanding of TBG is the mechanism of the charge carrier interaction, where some research groups proposed that there is a dominant role of the electron-phonon interaction (which is also considered as the emerging mechanism for the superconductivity in MATBG by some authors [16-21]), while the other groups showed evidences for the dominance of the electron-electron interaction [1,8,15], and, recently, new experiments demonstrated the dominance of the electon-magnon interaction [4,7,10]. Such a variety of the proposed interaction mechanisms in TBG reflects a large variety of physical effects which are simultaneously synergised to form the electronic state in these 2D materials.

In attempt to quantify these physical effects in metallic TBG superlattices here we proposed to generalize the Bloch-Grüneisen (BG) equation [22,23] which describes temperature dependent resistance in metallic compounds and which in its classical form can be written as:

$$R(T) = R_0 + A_1 \cdot T + \sum_n^{2,3,5} A_n \cdot \left(\frac{T}{T_\theta}\right)^n \cdot \int_0^{\frac{T_\theta}{T}} \frac{x^n}{(e^x-1)\cdot(1-e^{-x})} \cdot dx \tag{2}$$

where $R_0$ is the resistance at $T \rightarrow 0$ K, $T_\theta$ is the Debye temperature, $A_n$ is weighting parameters, and $n$ is the power-law exponent for which has theoretical integer values for certain single interaction mechanism [22-25]:

$$n = \begin{cases} 2 \;\; implies\ that\ the\ resistance\ is\ due\ to\ electron - electron\ interaction \\ 3 \;\; implies\ that\ the\ resistance\ is\ due\ to\ electron - magnon\ interaction \\ 5 \;\; implies\ that\ the\ resistance\ is\ due\ to\ electron - phonon\ interaction \end{cases} \tag{3}$$

However, it should be noted that entire BG integral (Eq. 2) has a linear limit for $n \rightarrow 1$:

$$\lim_{n\to 1} \left(\frac{T}{T_\theta}\right)^n \cdot \int_0^{\frac{T_\theta}{T}} \frac{x^n}{(e^x-1)\cdot(1-e^{-x})} \cdot dx \to \left(\frac{B}{T_\theta}\right) \cdot T \tag{4}$$

where $B$ is a constant, and from mathematics, the linear term in Eq. 1 can be also represented in for of integral part at $n \rightarrow 1$ with some multiplicative weighting factor $A_1$:

$$R(T) = R_0 + A_1 \cdot T + \sum_n^{\left(\lim_{n\to 1}\right),2,3,5} A_n \cdot \left(\frac{T}{T_\theta}\right)^n \cdot \int_0^{\frac{T_\theta}{T}} \frac{x^n}{(e^x-1)\cdot(1-e^{-x})} \cdot dx \tag{5}$$

It should be noted that Eq. 2 in its full form has been never applied for the analysis of experimental *R*(*T*) data, because the sum of strongly non-linear integrals over-parametrizes fitting procedure. Moreover, majority of all published works utilizes Eq. 1 where only electron-phonon integrand, i.e. $n = 5$, is included [26-28].

One of possible way to use an analytic power of Eq. 2 is to reduce the number of integrals to one, but use the power-law exponent *n* as a free-fitting parameter:

$$R(T) = R_0 + A_n \cdot \left(\frac{T}{T_\theta}\right)^n \cdot \int_0^{\frac{T_\theta}{T}} \frac{x^n}{(e^x-1)\cdot(1-e^{-x})} \cdot dx \tag{6}$$

If the fit of *R*(*T*) to Eq. 6 will converge, then deduced free-fitting parameter *n* should indicate main charge carrier scattering mechanism in given materials.

From the best author's knowledge, the approach to implement Eq. 6 has been reported only by Jiang *et al* [25] for $Sr_2Cr_3As_2O_2$, where the dominant role of the electron-magnon scattering (i.e. $n$ = 3.34 [25]), with insignificant part of the electron-phonon interaction ($n$ = 5) has been revealed.

It should be noted that a replacement full integrals in Eq. 2 or the integral in Eq. 6 by power law terms, $A_n \cdot T^n$, which has been implemented in several reports [29-31], cannot be accepted to be accurate approximation, as we show below herein.

It should be noted that linear dependence of $R(T)$ (or $n \rightarrow 1$ in terms of Eq. 6) in TBG has been proposed to be related to quasielastic scattering on acoustic phonon in MATBG [19] and, thus, deduced $n$-values in the range of $1 < n < 2$ have a clear interpretation as a sum of the electron-electron and elecrtron-quasielastic acoustic phonon interactions.

Here we implemented Eq. 6 to fit $R(T)$ data in TBG superlattices. First of all, we test the validity of Eq. 6 to be proper fitting tool for classical electron-phonon materials, including electron-phonon mediated superconductors, from which we chose $ReBe_{22}$ [26], as well as normal metal copper, and ferromagnetic iron and cobalt (for all pure metals raw $R(T)$ data was taken from classical paper by White and Woods [32]), as well as for highly-compressed ε-phase of iron, which exhibits the superconducting state (for which raw $R(T)$ data were reported by Shimizu *et al* [33] and by Jaccard *et al* [29]. In Fig. 1 we show $R(T)$ data and data fits for these materials (it should be noted that fits for superconducting $ReBe_{22}$ and ε-Fe iron was performed by recently proposed equation [34]:

$$R(T) = R_0 + \theta(T_c^{onset} - T) \cdot \left( \frac{R_{norm}}{\left( I_0\left( F \cdot \left(1 - \frac{T}{T_c^{onset}}\right)^{3/2} \right) \right)^2} \right) + \theta(T - T_c^{onset}) \cdot \left( R_{norm} + A \cdot \left( \left(\frac{T}{T_\theta}\right)^n \cdot \int_0^{\frac{T_\theta}{T}} \frac{x^n}{(e^x - 1)\cdot(1 - e^{-x})} \cdot dx - \left(\frac{T_c^{onset}}{T_\theta}\right)^n \cdot \int_0^{\frac{T_\theta}{T_c^{onset}}} \frac{x^n}{(e^x - 1)\cdot(1 - e^{-x})} \cdot dx \right) \right) \quad (7)$$

but where now we changed the $n$-value to be a free-fitting parameter, and where $T_c^{onset}$ is free-fitting parameter of the onset of superconducting transition, $R_{norm}$ is the sample resistance at the onset of the transition, $\theta(x)$ is the Heaviside step function, $I_0(x)$ is the zero-order modified Bessel function of the first kind and $F$ is a free-fitting dimensionless parameter.

It can be seen, that expected $n = 5$ has been revealed for electron-phonon mediated $ReBe_{22}$, and reasonable value of $n = 4.3 \pm 0.3$ was revealed for pure copper, where the limiting number of raw experimental $R(T)$ was the most likely a reason for slightly lower than $n = 5$ value. However, we keep the use of $R(T)$ datasets reported by White and Woods [32], because they reported $R(T)$ data for γ-Fe and Co measured by the same experimental routine and because the majority of $R(T)$ data reported for MATBG also contain limited (by employing a wide temperature step for measurements) $R(T)$ datasets.

Our analysis by Eq. 6 reveals that γ-Fe, which should have $n = 3$ [24,25,32], exhibits $n = 2.9 \pm 0.1$ which is an excellent demonstration for the applicability of Eq. 6 to the analysis. Ferromagnetic cobalt has $n = 2.2 \pm 0.1$, which reflects a well-established fact that electron-electron interaction in this element plays significant role [32].

Another interesting result was obtained for hexagonal-close-packed highly-compressed iron, ε-Fe. This ε-Fe phase plays crucial role in the Earth geology [31], because its electrical conductivity, $\rho$, directly links with the heat transfer in the Earth crust due to the Wiedemann-Franz law:

$$k(T,P) = \frac{L \cdot T}{\rho(T,P)} \tag{8}$$

where $k$ is the thermal conductivity and $L = 2.44 \cdot 10^{-8}$ WΩK$^{-2}$ is the Lorenz number, and $P$ is the pressure (details can be found in Ref. 31).

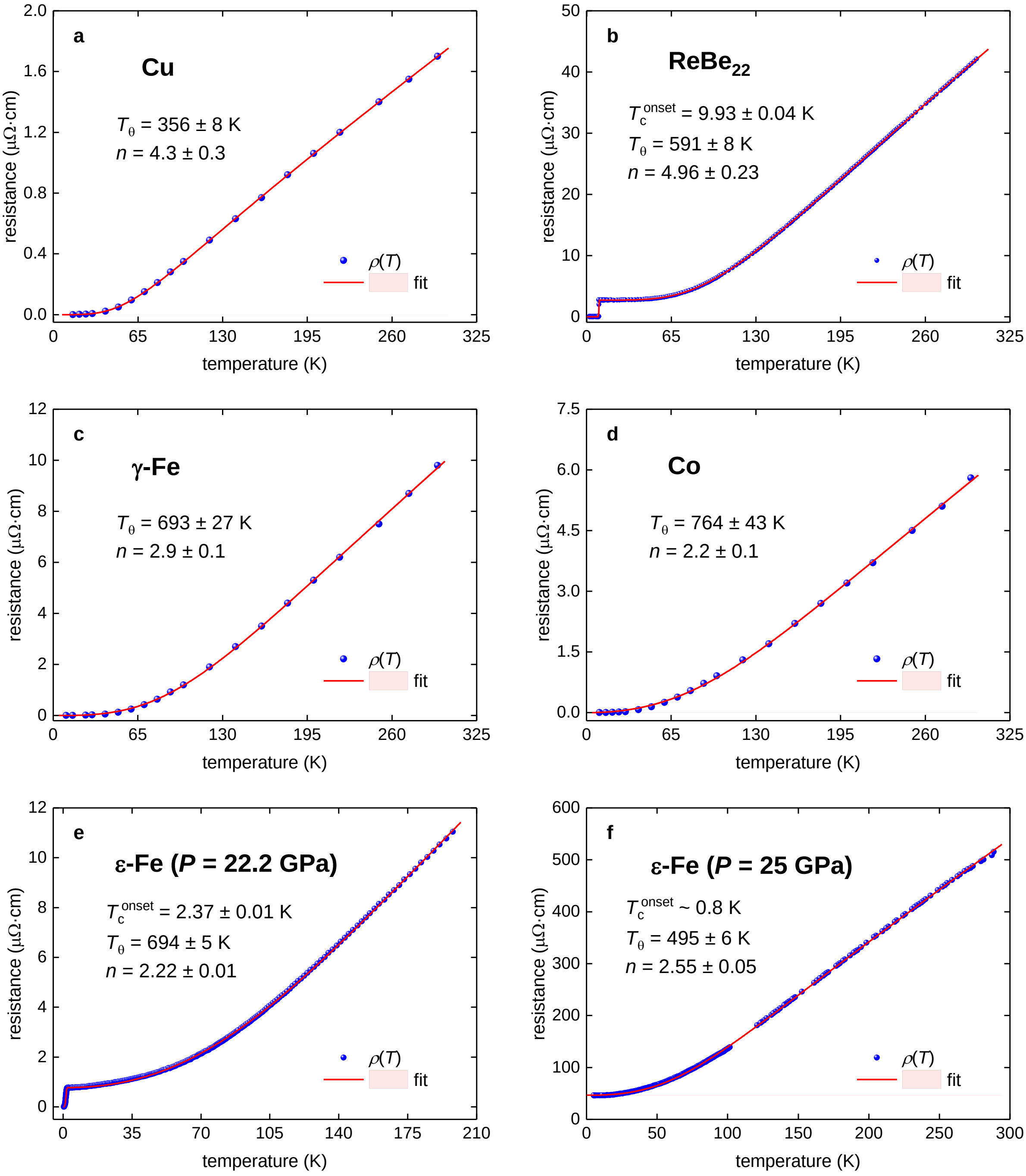

**Figure 1.** $\rho(T)$ data and fits to generalized Bloch-Grüneisen (BG) equation (Eqs. 5,6) for pure Cu (a), $ReBe_{22}$ (b), pure ferromagnetic γ-Fe (c), pure ferromagnetic Co (d), and pure non-ferromagnetic highly-compressed ε-Fe (e,f). Raw data reported in Refs. 26,29,32,33. Red is the fitting curve, and 95% confidence bars are shown by a pink shaded area. Goofiness of fit for all datasets are better than $R$ = 0.9990.

From the best knowledge of the author's, to date, experimental $\rho(T,P)$ data was fitted only to an approximant function of Eq. 6 which has a form of a power-law, where $\alpha$ and $\beta$ are free-fitting parameters:

$$\rho(T) = \beta \cdot T^n. \quad (9)$$

In result, reported α values are within an extremely wide range of $n = 1.5 - 5.9$, and, moreover, we found herein that the approach to use Eq. 9 leads to wrong $n$-values. Truly, in Fig. 1,e we show the fit to Eq. 6, which reveals $n = 2.22 \pm 0.01$ for which, by the employing the same $\rho(T)$ dataset, and the use of Eq. 9, Jaccard *et al* [29] reported $n = 1.67$. If our value of $n = 2.22 \pm 0.01$ shows that electric charge carriers in ε-Fe phase exhibits two scattering mechanisms (i.e., mainly the electron-electron interaction ($n$ = 2) with some weighting part of the electron-magnon interaction ($n$ = 3)), the interpretation for $n = 1.67$ reported by Jaccard *et al* [29] cannot be founded, because $n \rightarrow 1$ case is only applicable for MATBG superlattices [19], and $n$-value below 2 are simply prohibited for elemental metals, because there is no physical interpretation for such values.

It is important to note, that there is a nice correlation between $n$-values and superconducting transition temperatures, Tc, in ε-Fe phase too. If for $n = 2.55 \pm 0.05$ (which implied a significant electron-magnon interaction) the full resistive transition does not occurs (and where the only 10% drop in resistance observed, with the onset of transition temperature, $T_c^{onset} \sim 1\ K$), for ε-Fe sample, for which $n = 2.22 \pm 0.01$ was revealed the full resistive transition was observed with $T_c^{onset} = 2.37 \pm 0.01\ K$. This result has a clear interpretation that the suppression of the electron-magnon interaction causes the formation of more robust superconducting condensate.

Now, we turn to the analysis of TBG superlattices. First we analysed experimental $R(T)$ curves for Moiré superlattice in single layer graphene on hBN single crystal (SLG/hBN superlattice) reported by Wallbank *et al* [8], where the Moiré superlattice constant, λ, has been changed in the range of λ = 11.2 – 15.1. Fits to Eq. 6 are shown in Fig. 2 and summarized results in Fig. 3. It can be seen that, in overall, our analysis confirms the result reported by Wallbank *et al* [8], that the electron-electron interaction is dominant in these Moiré 2D supelattices. However, our analysis shows a smooth and near-linear dependence of $n$-value and the Debye temperature, $T_\theta$, on the Moiré superlattice constant, λ (Fig. 3).

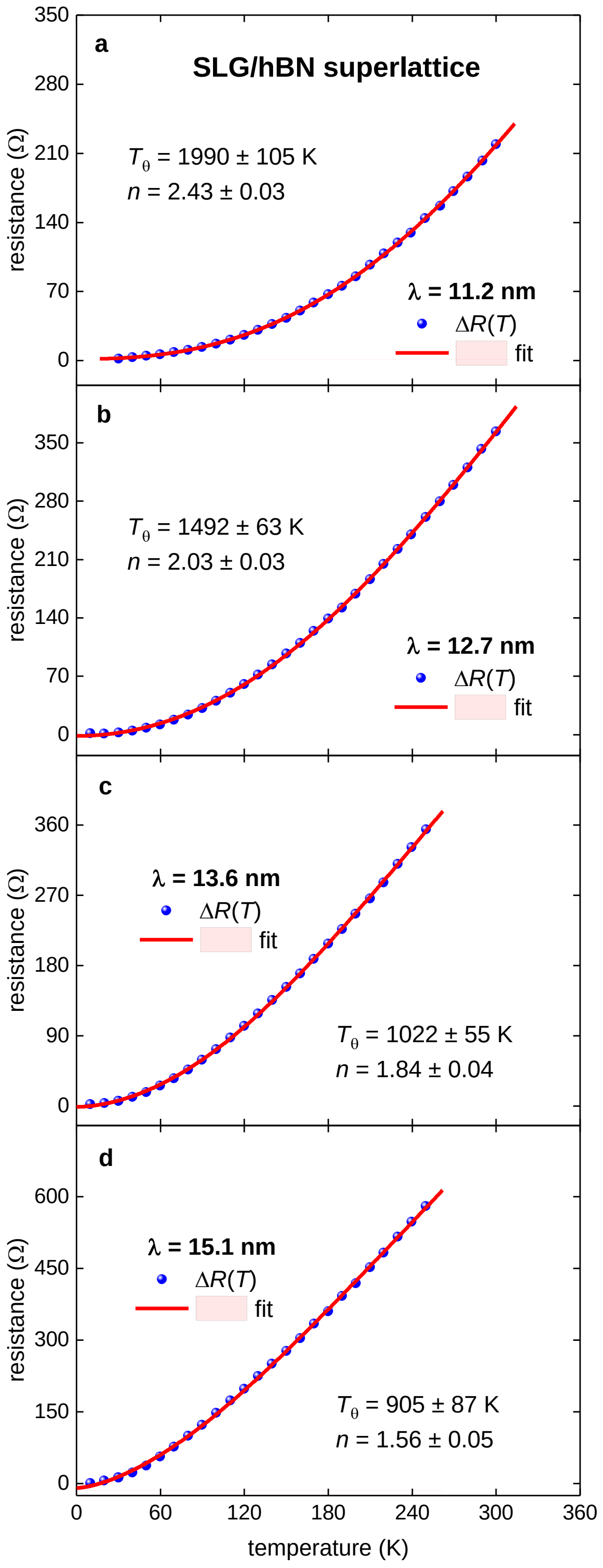

**Figure 2.** $R(T)$ data for Moiré SLG/hBN superlattices (raw data reported by Wallbank *et al* [23]) and fit to Eq. 6 for (a) λ = 11.2 nm; (b) λ = 12.7 nm; (c) λ = 13.6 nm, (d) and 15.1 nm. Red are the fitting curves, 95% confidence bars are shown by a pink shaded area. Goofiness of fit for all plots was better than $R = 0.9997$.

As we mentioned above, the power-law exponent $n$ towards lower than 2 values in graphene/hBN superlattices (Fig. 2(e)) has now clear interpretation that $n < 2$ values represent the transition from pure electro-electron interaction (for which the characteristic value is $n = 2$) to some intermediate state exhibited a synergetic sum of the electron-electron (*e-e*) and the electron-quasielastic acoustic phonon (*e-qaph*) interactions.

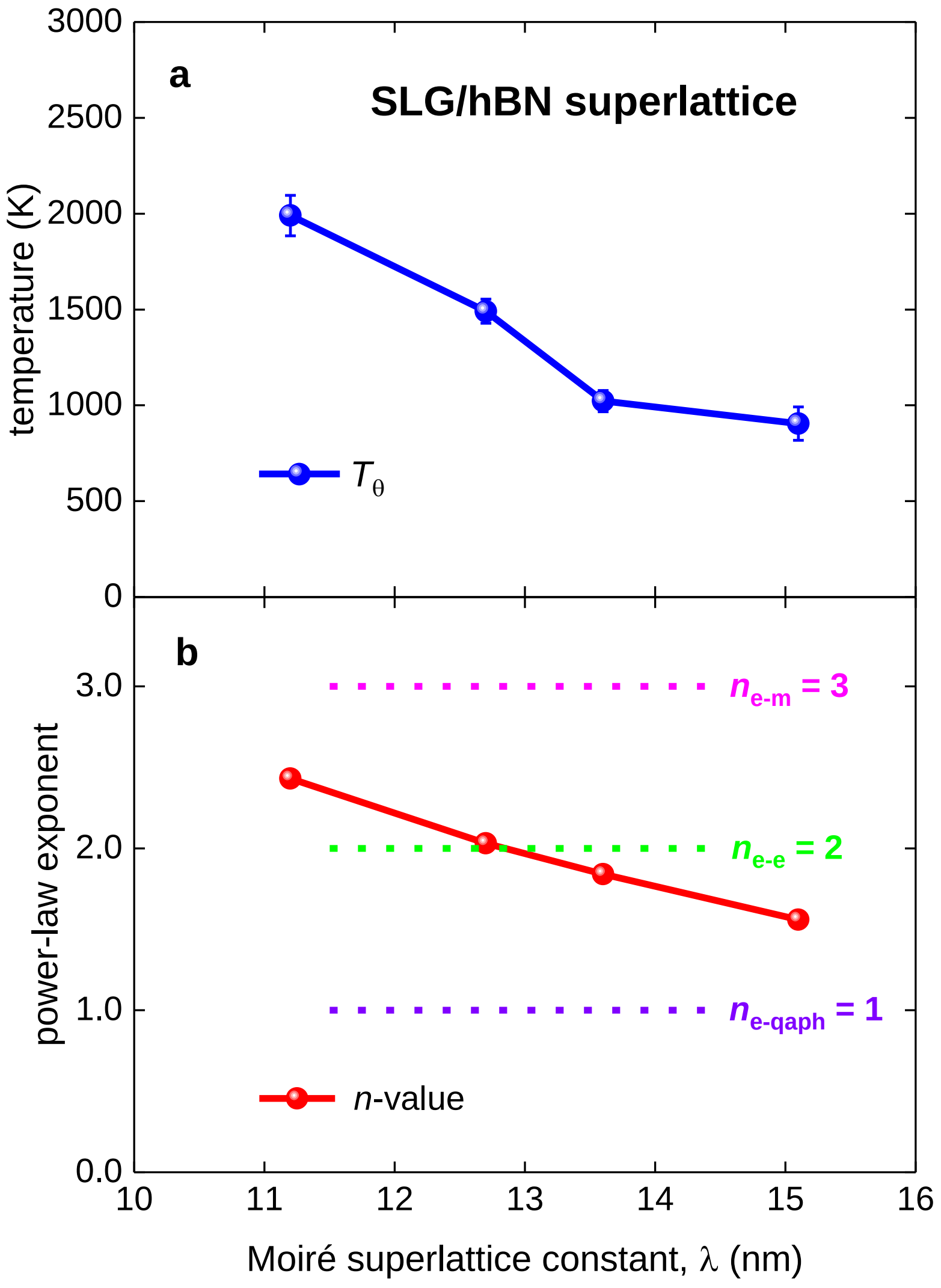

**Figure 3.** Summarized results for Moiré SLG/hBN superlattices (raw data reported by Wallbank *et al* [23]). ***a*** – deduced Debye temperature; ***b*** – deduced $n$-value in Eq. 5. Characteristic values for the quasielastic electron-acoustic phonon interaction ($n_{e\text{-}qaph} = 1$), the electron-electron interaction ($n_{e\text{-}e} = 2$), and the electron-magnon interaction ($n_{e\text{-}m} = 3$) are shown. Error bars for $n$-value are less than the balls size.

This mixed state characterized by 1 < $n$ < 2 has been also revealed in the metallic twisted bilayer graphene stabilized by $WSe_2$ (for which raw $R(T)$ data was reported by Arora *et al* [9]). In Fig. 3 we show $R(T)$ data and fits for samples with twisted angles $\theta = 0.87°$ (filling factor ν = +1, deduced $n = 1.52 \pm 0.05$, $T_\theta = 47 \pm 9\ K$) and 0.97° (filling ν = -1, deduced $n = 1.75 \pm 0.09$, $T_\theta = 13.0 \pm 0.8\ K$).

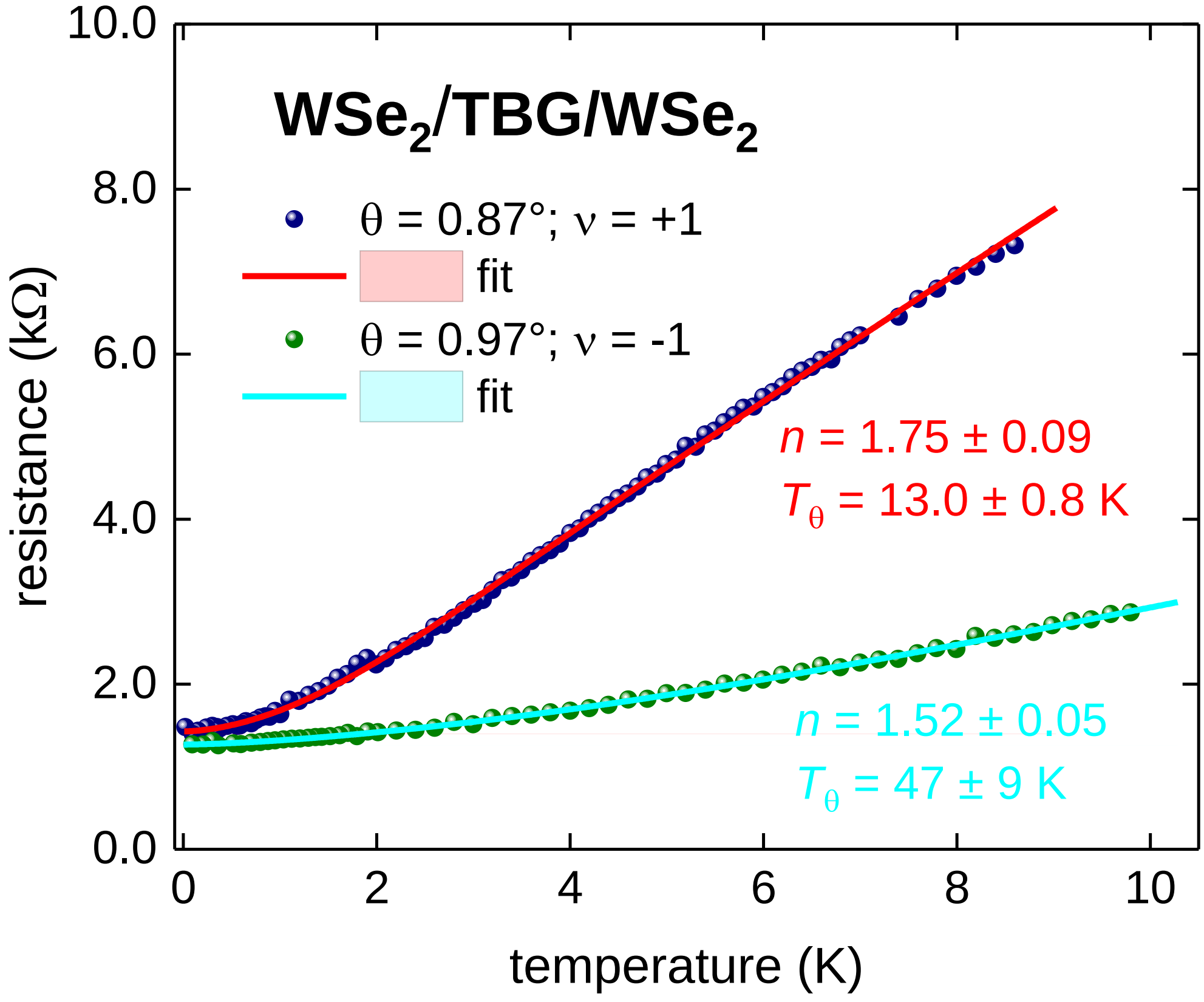

**Figure 4.** $R(T)$ data for metallic bilayer Moiré graphene superlattices (raw data reported by Arora *et al* [9]) and fit to Eq. 5 for $\theta = 0.87°$ and $\theta = 0.97°$. 95% confidence bars are shown for both fitting curves. Goodness of fits for both plots was better than $R = 0.998$.

The majority of experimental studies in TBG superlattices have been performed for the $R(T)$ dependences on the charge carrier density. In this work, we performed the analysis for the metallic states of TBG system which exhibits the twisted angle of $\theta = 2.02°$ (for which the raw experimental $R(T)$ was reported by Polshyn *et al* [6]). Data was analysed in full range of the charge carrier density of $p \leq \pm|6.71| \cdot 10^{12}\ cm^{-2}$. We presented herein results for $R(T)$ data fits which was undertaken at $T \leq 192.5\ K$. Representative fittings where $n$-value is reaching the characteristic values of $n$ = 2, $n$ = 3, as well as a low value of $n$ = 1.4 and the highest value of $n$ = 4.7 are shown in Figs. 5,6. We do not perform fit to Eq. 6 for $R(T)$ curves measured at very low charge carrier density, where low-temperature upturn in the $R(T)$ curve was observed.

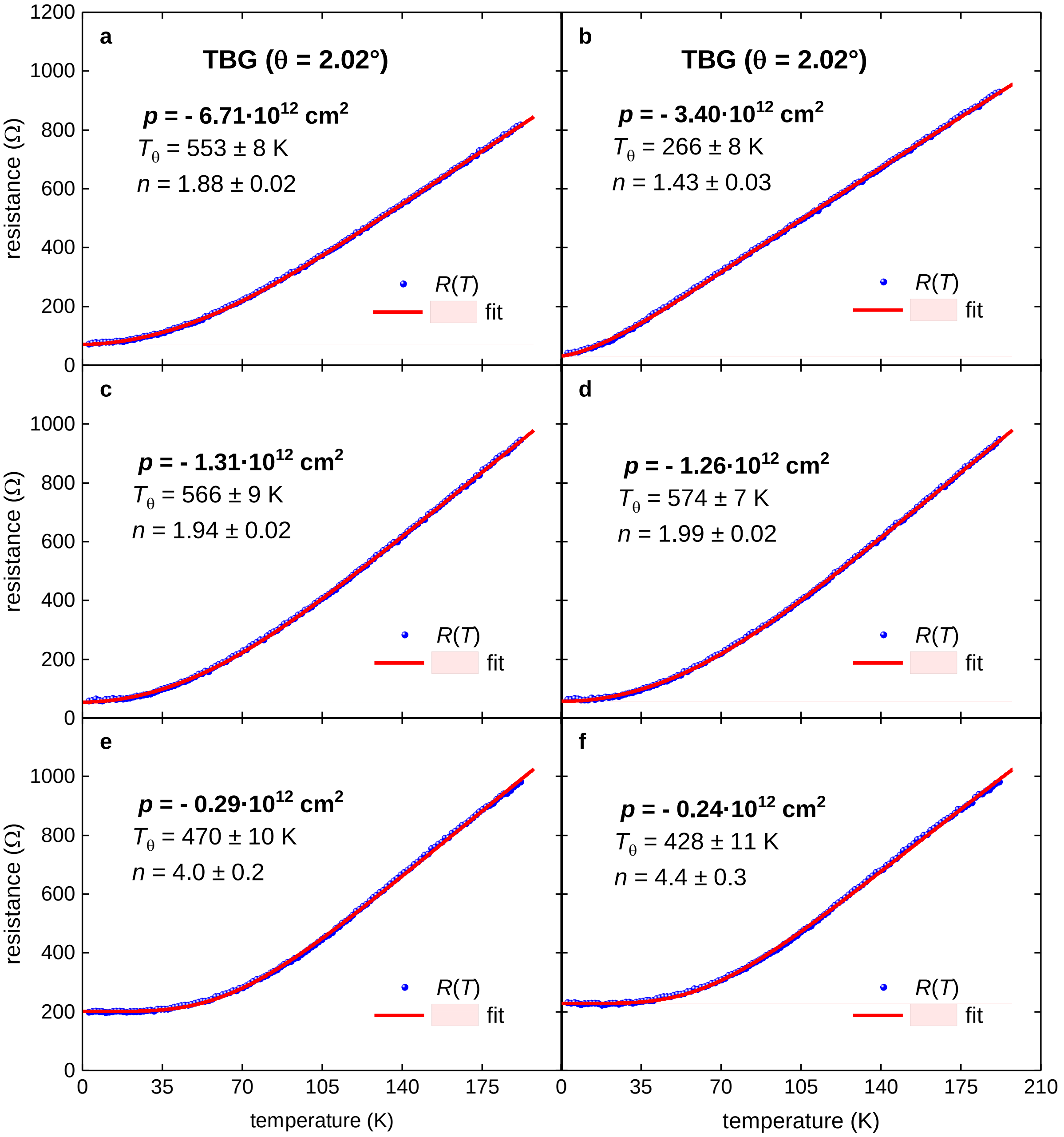

**Figure 5.** $R(T)$ data and data fit to Eq. 5 for metallic TBG superlattice on hole side with $\theta$ = 2.02° (raw $R(T)$ data was reported by Polshyn *et al* [6]). Red are the fitting curves, 95% confidence bars are shown by a pink shaded area. Goofiness of fit for both plots was better than $R$ = 0.9990.

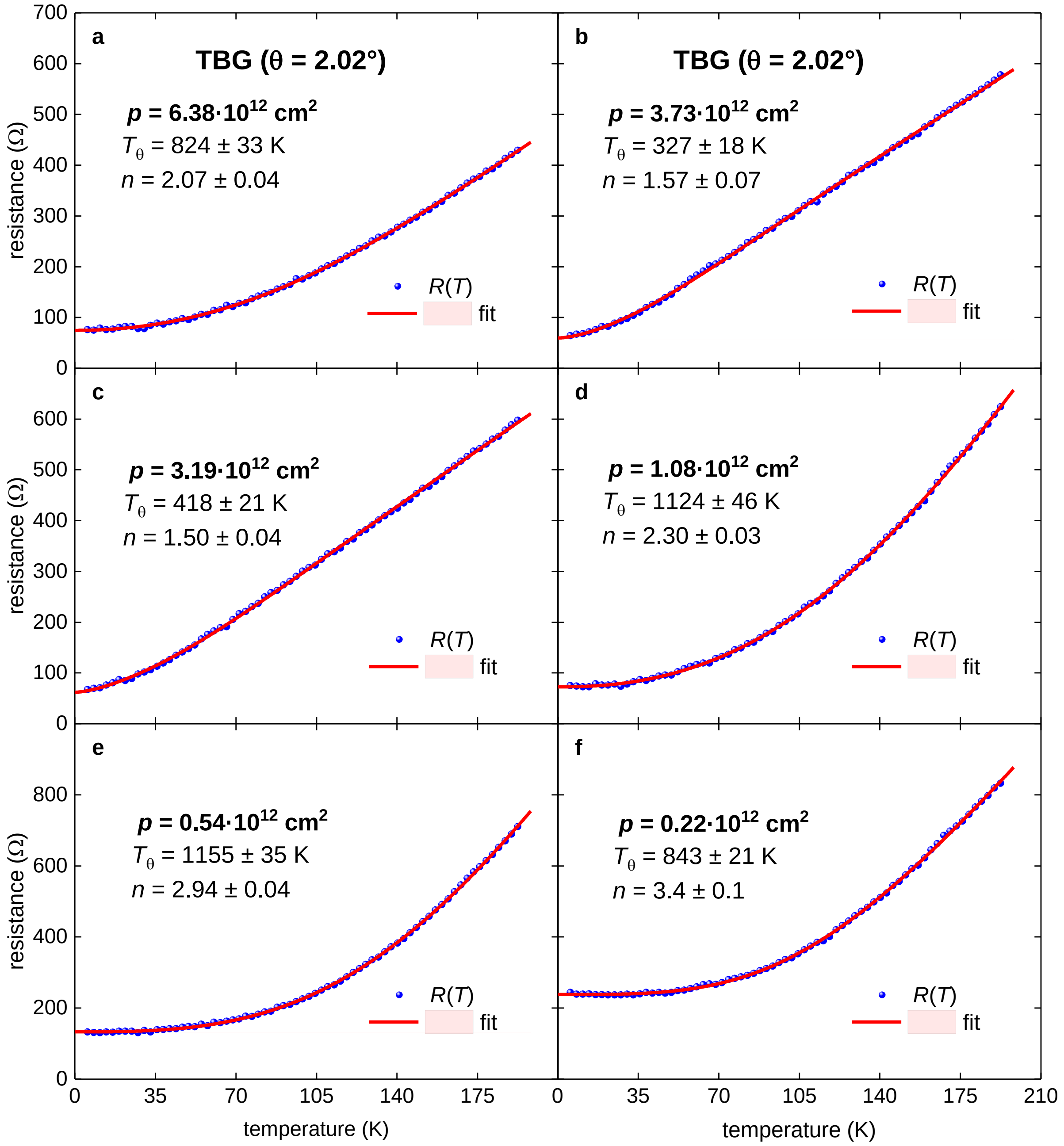

**Figure 6.** $R(T)$ data and data fit to Eq. 6 for metallic TBG superlattice on electron side with $\theta$ = 2.02° (raw $R(T)$ data was reported by Polshyn *et al* [6]). Red are the fitting curves, 95% confidence bars are shown by a pink shaded area. Goofiness of fit for both plots was better than $R$ = 0.9990.

Summarized result for the Debye temperature, $T_\theta(p)$, and the power-law exponent, $n(p)$, vs the charge carrier density are shown in Fig. 7. There are several important findings:

1. Classical electron-phonon interaction (with $n > 3.5$) can be observed at lowest charge carrier concentration in a very narrow concentration range, $-0.39 \cdot 10^{12}\ cm^{-2} <$ $p < -0.39 \cdot 10^{12}\ cm^{-2}$. In this concentration range we where we skipped from the analysis several $R(T)$ curves measured at very low $p$, which exhibit upturn in $R(T)$ at $T$ < 20 K.

2. The dominant role of the electron-magnon interaction (2.5 < $n$ < 3.5) has been revealed at low charge carrier concentration, $|0.4| \cdot 10^{12}\ cm^{-2} < p < |1.0| \cdot 10^{12}\ cm^{-2}$.

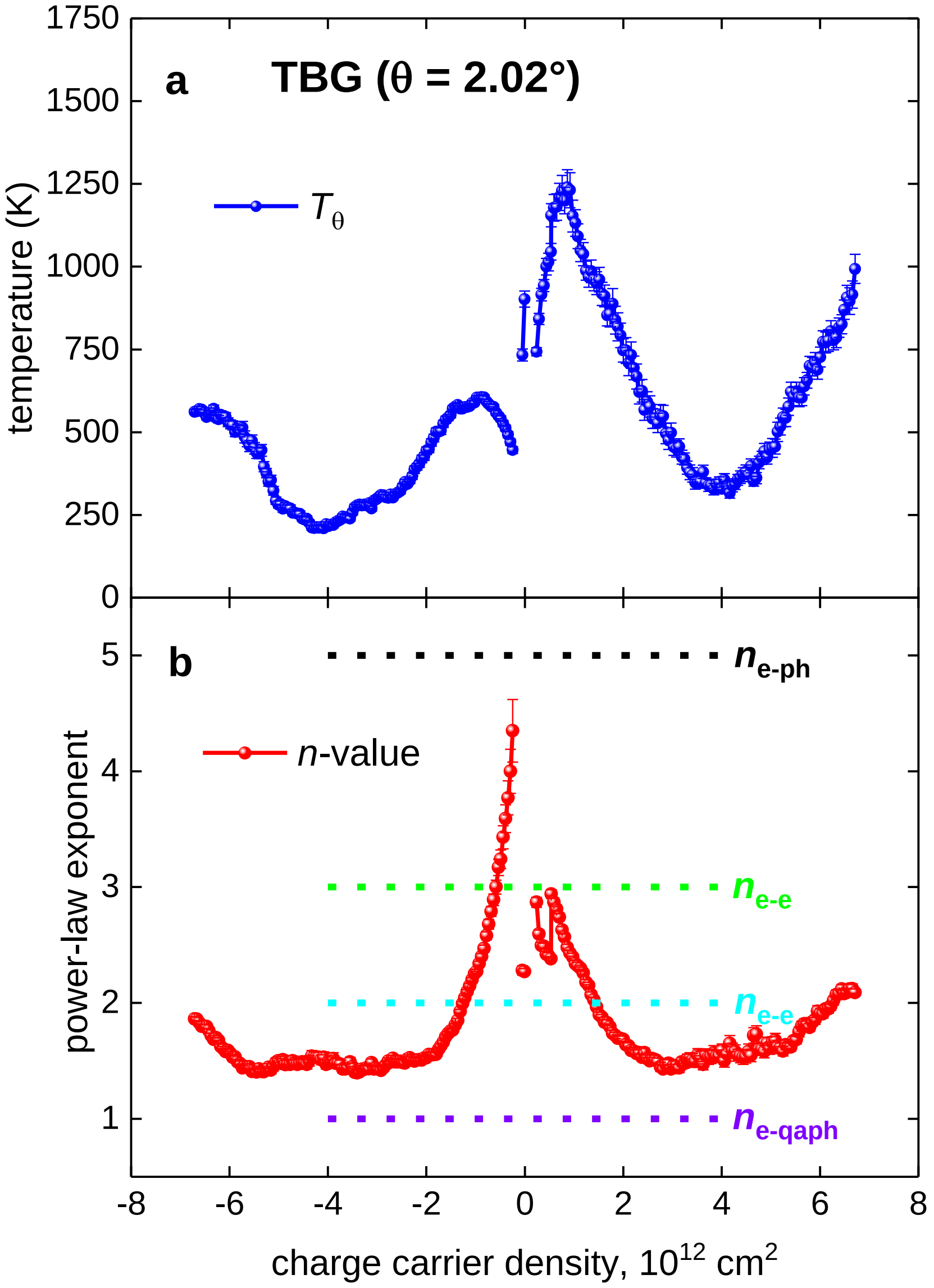

**Figure 7.** Summarized results for MATBG superlattice with $\theta$ = 2.02°. ***a*** – deduced Debye temperature; ***b*** – deduced $n$-value. Characteristic values for the quasielastic electron-acoustic phonon interaction ($n_{e\text{-}qaph}$ = 1), the electron-electron interaction ($n_{e\text{-}e}$ = 2), and the electron-magnon interaction ($n_{e\text{-}m}$ = 3), and the electron-phonon interaction ($n_{e\text{-}ph}$ = 5) are shown. Raw $R(T)$ data was reported by Polshyn *et al* [6].

3. In a wide range of doping, $|1.0| \cdot 10^{12} \lesssim p \lesssim |6| \cdot 10^{12}\ cm^{-2}$, the interaction is belonging a sum of the electron-electron and the electron-quasielastic acoustic phonon interactions.
4. And only at high charge carrier density, $p > |5.5| \cdot 10^{12}\ cm^{-2}$, the electron-electron interaction overcomes the others interaction mechanisms, and power-law exponent towards $n = 2$, while the doping is increasing.

In summary, in this paper we aim to propose an approach to quantify the charge carrier integration in metallic materials by generalizing Bloch-Grüneisen equation, where power-law exponent is a free-fitting parameter. In particular case of twisted bilayer graphene superlattices we show that the interaction mechanism can be smoothly transformed from one to another by a variation of either the Moiré superlattice constant, λ, or the charge carrier concentration. We also show that generalized Bloch-Grüneisen equation can be an instructive tool to study different topics in natural science, including the Earth geology.

**Acknowledgement**

The author thanks Dr. Gregory Polshyn and Prof. Andrea F. Young (University of California at Santa Barbara) for kind provide an extended experimental dataset for TBG superlattices analysed in this work. The author thanks financial support provided by the Ministry of Science and Higher Education of Russia (theme “Pressure” No. AAAA-A18-118020190104-3) and by Act 211 Government of the Russian Federation, contract No. 02.A03.21.0006.